\documentclass[10pt,a4paper]{article}

\usepackage{bm,graphicx}

\def\Extr{\mathop{\rm Extr}}

\begin{document}
\bibliographystyle{unsrt}

\title{Unsupervised learning of binary vectors: a Gaussian scenario}
\author{Mauro Copelli\thanks{eletronic address: {\tt
mauro@hypatia.ucsd.edu}} \\ 
\small Department of Chemistry and Biochemistry 0340, 
\small University of California San Diego\\
\small La Jolla, CA 92093-0340, USA \and 
Christian Van den Broeck\thanks{eletronic
address: {\tt christian.vandenbroeck@luc.ac.be}} \\
\small Limburgs Universitair Centrum, B-3590 Diepenbeek, Belgium}

\date{\today}

\maketitle

\begin{abstract}
We study a model of unsupervised learning where the real-valued data
vectors are isotropically distributed, except for a single symmetry
breaking binary direction $\bm{B}\in\{-1,+1\}^{N}$, onto which the
projections have a Gaussian distribution.  We show that a candidate
vector $\bm{J}$ undergoing Gibbs learning in this discrete space,
approaches the perfect match $\bm{J}=\bm{B}$ exponentially.  Besides
the second order ``retarded learning'' phase transition for unbiased
distributions, we show that first order transitions can also occur. 
Extending the known result that the center of mass of the Gibbs
ensemble has Bayes-optimal performance, we show that taking the sign
of the components of this vector (clipping) leads to the vector with optimal
performance in the binary space.  These upper bounds
are shown generally not to
be saturated with the technique of transforming the components of a
special continuous vector, except in asymptotic limits and in a
special linear case.  Simulations are presented which are in excellent
agreement with the theoretical results.
\end{abstract}

%\maketitle must follow title, authors, abstract and \pacs

\section{Introduction}
\label{intro}

Since the introduction of the Ising spin model, the study of models
with discrete degrees of freedom has become a core activity in
statistical mechanics.  When combined with disorder, such models often
have interesting connections to problems of computational complexity,
to learning theory or to open problems in statistics.  Discreteness
and disorder introduce intrinsic difficulties, and exactly solvable
models are rare.  The main purpose of this paper is to present a
discrete model with disorder which can be solved in full detail.  The
model is most naturally presented as an unsupervised learning problem,
and we briefly review the connection with the existing literature.

The goal of unsupervised learning is finding structure in
high-dimensional data.  In one of the simplest parametric models
introduced in the
literature~\cite{Biehl93a,Biehl94b,Watkin94a,Reimann96a,%
Reimann96b,VandenBroeck96,Buhot98,Gordon98,Herschkowitz99},
$N$-dimensional independently drawn data vectors
$D=\{\bm{\xi}^{\mu}\}$, $\mu=1,\ldots,\alpha N$, are uniformly
distributed, except for a single symmetry-breaking direction $\bm{B}$. 
If we assume that all the relevant probability distributions are
known, the aim of learning is to construct an estimate vector $\bm{J}$
of the true direction $\bm{B}$.

Previous studies of this model focused on the case where $\bm{B}$ is
constrained to have a constant size, being otherwise equiprobably
sampled from the $N$-sphere.  This so-called {\it spherical case\/} is
associated with a spherical {\it prior distribution\/} 
%(hereafter referred to simply as {\it prior\/}) 
$P_{s}(\bm{B})\sim \delta(\bm{B}\cdot\bm{B}-N)$.  The focus of the
present paper, however, is on binary (or Ising) vectors.  In this
case, $\bm{B}$ is known to have binary components only,
$B_{j}\in\{-1,+1\}$, $j=1,\ldots,N$.  This extra knowledge is taken
into account by assigning a binary prior distribution

\begin{equation}
\label{binprior}
P_{b}(\bm{B}) = \prod_{j=1}^{N}\left[
\frac{1}{2}\delta(B_{j}-1) + 
\frac{1}{2}\delta(B_{j}+1)
\right]
\end{equation}
to the preferential direction.

In this framework, a Gaussian scenario was introduced
in~\cite{Reimann96b} as a kind of minimal model, allowing the
calculations to be much simplified and the spherical case to be solved
exactly.  In this model, the components of $\bm{\xi}$ perpendicular
to $\bm{B}$ are assumed to be independent Gaussian distributed variables with
zero mean and unit variance,
i.e. $P(b^{\prime}\equiv
\bm{B^\prime}\cdot\bm{\xi}/\sqrt{N}) =
\exp(-{b^{\prime}}^2/2)/\sqrt{2\pi}$, where
$\bm{B^\prime}\cdot\bm{B}/N = 0$.  
The distribution of the component
$b\equiv \bm{B}\cdot\bm{\xi}/\sqrt{N}$ parallel to $\bm{B}$, on the
other hand, can be chosen at will, and in the Gaussian scenario it is
completely determined by the mean $B$ and variance $1-A$:

\begin{eqnarray}
\label{pofb}
{\cal P}(b) & = & \frac{\cal N}{\sqrt{2\pi}} \exp\left\{ - 
\frac{b^{2}}{2} - U(b) \right\}\; ,  \\
\label{UGauss}
U(b) & = & \frac{A}{2(1-A)}b^{2} - \frac{B}{1-A}b\; ,
\end{eqnarray}
where ${\cal N} = [\int{\cal D}b\, \exp -U(b)]^{-1}
%=\exp[-B^{2}/(2(1-A))]/\sqrt{1-A}
$ is a normalization constant and ${\cal D}b\equiv db\,
\exp[-b^{2}/2]/\sqrt{2\pi}$.

In comparison with the spherical case, the binary case presents
several extra difficulties, which motivates the study of this simple
model.  The main question to be addressed in this work is: given the
$\alpha N$ data vectors (also called patterns) and the knowledge of
the probability distributions, what is the best estimate $\bm{J}$ one
can construct to approximate $\bm{B}$?  The answer, cast in the
framework of Bayesian inference, depends on whether $\bm{J}$ is
allowed to have continuous components or, conversely, is required to
be a binary vector.  We also address the problem of whether these
upper bounds can be simply attained, by first obtaining a continuous
vector via minimization of a potential and then transforming its
components.

The results of the replica calculation for this problem are briefly
reviewed in section~\ref{general}.  Section~\ref{Gibbs} discusses the
special case of Gibbs learning, for which simulations have been
performed.  In section~\ref{Bayes} we review the reasoning leading to
the Bayesian bound in the continuous as well as the binary space, with
simulations compared to the theoretical results.  A simple strategy
which attempts to saturate these upper bounds is studied in
section~\ref{transform}, while our conclusions are presented in
section~\ref{conclusions}.

\section{Unsupervised learning}
\label{general}

In order to obtain a good candidate vector $\bm{J}$, we construct a
cost function of the form ${\cal H} = \sum_{\mu}^{\alpha
N}V(\lambda_{\mu})$, where $\lambda_{\mu}\equiv
\bm{J}\cdot\bm{\xi}^{\mu}/\sqrt{N}$.  In the Gaussian scenario, the
{\em potential\/} $V$ has a quadratic form,

\begin{equation}
V(\lambda) = \frac{c}{2}\lambda^{2} - d\lambda\; .
\end{equation}
Learning is defined as sampling $\bm{J}$ from the Boltzmann
distribution with temperature $T=1/\beta$

\begin{equation}
\label{PJgivenD}
P(\bm{J}|D) = \frac{P(\bm{J})}{Z}\exp -\beta\sum_{\mu}^{\alpha
N}V(\lambda_{\mu})\; ,
\end{equation}
where $Z(D) = \int d\bm{J}\, P(\bm{J}) \exp -\beta{\cal H}$ is the
normalization constant and the measure $P(\bm{J})$ is used to enforce
either a binary ($P(\bm{J})=P_{b}(\bm{J})$) or a spherical
($P(\bm{J})=P_{s}(\bm{J})$) constraint on $\bm{J}$.  While the
spherical case has been dealt with in~\cite{Reimann96b}, we focus now
on the case where the candidate vectors have binary components.  The
thermodynamic properties of such a system can be read from the free
energy $f \equiv -(1/\beta N)\ln Z$.  In the thermodynamic limit
$N\to\infty$, $f$ becomes self-averaging, $\left<f(D)\right>_{D} = f$,
and can be calculated via the replica trick.  This by now standard
calculation will not be reproduced here, only the results are quoted:
for a replica symmetric {\it ansatz\/}, the quadratic forms of the
Gaussian scenario allow the calculations to be performed exactly, and
the free energy reads

%\begin{widetext}
\begin{eqnarray}
\label{freebin}
f & = & \Extr_{q,R,\hat{q},\hat{R}}\Bigg\{ \frac{(1-q)\hat{q}}{2\beta}
+ \frac{\hat{R}R}{\beta}
%\nonumber \\ & \ &
-\frac{1}{\beta}\int{\cal D}z\,\ln\cosh\left(z\sqrt{\hat{q}}+\hat{R}\right) 
+\frac{\alpha}{2\beta}\ln[1+c\beta(1-q)]
\nonumber \\  & \ & 
+\alpha
\left[
\frac{
c [q-R^{2}(A-B^{2})] - 2BRd - d^{2}\beta(1-q)
}{
2[1+c\beta(1-q)]
}
\right] \Bigg\}\; .
\end{eqnarray}
%\end{widetext}
The above order parameters are also self-averaging quantities and can
be interpreted as follows: $R = \bm{B}\cdot\bm{J}/N$ measures the
alignment between a typical (binary) sample of eq.~\ref{PJgivenD} and
the preferential direction, its absolute value as a function of
$\alpha$ being hereafter used to account for the performance of a
given potential $V$; $q = \bm{J}\cdot\bm{J}^{\prime}/N$ is the mutual
overlap between two different samples of eq.~\ref{PJgivenD}, while
$\hat{R}$ and $\hat{q}$ are the associated conjugate parameters.  The
equilibrium values of the four variables are determined by the
solution of the saddle point equations which arise from the extremum
operator in eq.~\ref{freebin}.

\section{Gibbs learning}
\label{Gibbs}

Gibbs learning arises as a particular but very important case in this
general framework.  In order to define it properly, we first recall
the Bayes inversion formula 

\begin{equation}
\label{posterior}
p(\bm{B}|D) \equiv \frac{P(D|\bm{B})P_{b}(\bm{B})}{P(D)}\; .
\end{equation}
The {\em posterior distribution\/} $p(\bm{B}|D)$ expresses the
knowledge about $\bm{B}$ which is gained after the presentation of the
data.  Replacing $\bm{B}$ with $\bm{J}$ in this formula gives the
probability density that $\bm{J}$ is the ``true'' direction $\bm{B}$,
given the data vectors.  Note that the binary prior in
eq.~\ref{posterior} constrains the acceptable candidates $\bm{J}$ to
the corners of the $N$-hypercube, i.e. $\bm{J}\in\{-1,+1\}^{N}$. 
Making use of eq.~\ref{pofb}, one rewrites

\begin{eqnarray}
\label{gibbslearning}
p(\bm{J}|D)  & \propto & 
P_{b}(\bm{J}) \prod_{\mu}^{p}
\exp -U\left(\bm{J}\cdot\bm{\xi}^{\mu}/\sqrt{N}\right)
\; ,
\end{eqnarray}
apart from a normalization constant.  Gibbs learning is defined as
sampling from distribution~\ref{gibbslearning}.

A comparison with eq.~\ref{PJgivenD} shows that the thermodynamics of
such a process is obtained by setting
$\beta V=U$~\cite{Watkin94a,Reimann96a}. Upon substitution of
$\beta=1$, $c = A/(1-A)$ and $d=B/(1-A)$ in eq.~\ref{freebin},
one finds that the extremum of the corresponding free-energy is
reached for $q_{G}=R_{G}$ and $\hat{q}_{G}=\hat{R}_{G}$, where 
the subscript $G$ will hereafter be used to denote results 
from Gibbs learning. The equalities reflect the symmetric role
played by $\bm{J}$ and $\bm{B}$ in Gibbs learning, a property
which has been previously noted in several publications (see
e.g.~\cite{Gyorgyi90} and~\cite{Opper94a}, among others). The four 
original saddle point equations are then effectively reduced 
to a single one:

\begin{equation}
\label{spRG}
R_{G} = F_{B}^{2}\left({\cal F}\left(\sqrt{R_{G}}\right)\right)\; ,
\end{equation}
where 

\begin{equation}
\label{FB}
F_{B}(x) \equiv \sqrt{
\int{\cal D}z\,\tanh(zx+x^{2})
}
\end{equation}
is a function coming from the entropic term of the free energy, while

\begin{equation}
\label{GL:gauss:sp:sp}
{\cal F}^{2}\left(\sqrt{R_{G}}\right) = 
\alpha\left[
\frac{B^{2} + AR_{G}(A-B^{2})}{1-AR_{G}}
\right]\; .
\end{equation}
The solution of eq.~\ref{spRG} also determines the value of the
conjugate parameter $\hat{R}_{G}$:

\begin{equation}
\label{spRGhat}	
\hat{R}_{G} = {\cal F}^{2}\left(\sqrt{R_{G}}\right)\; .
\end{equation}

In order to check that the replica symmetric {\it ansatz\/} is
correct, we also study the entropy $s \equiv
\beta^{2}\frac{\partial}{\partial\beta}[f - (\ln 2)/\beta]$, which for
Gibbs learning reads

\begin{eqnarray}
\label{sG}
s_{G} & = & -\frac{(1+R_{G})\hat{R}_{G}}{2} +
\int {\cal D}z\,\ln 2\cosh\left(z\sqrt{\hat{R}_{G}} + 
\hat{R}_{G}\right) \nonumber \\
& \ & -\frac{\alpha}{2} \left[ 
\ln\left(\frac{ 1-AR_{G}}{1-A}\right) + 
(B^{2}-A)(1-R_{G}) \right] \; .
\end{eqnarray}
On physical grounds, this quantity should always remain positive. 
Additionally, by relating $s_{G}$ with the mutual information $i$ per
degree of freedom between the data $D$ and the preferential direction
$\bm{B}$, Herschkowitz and Nadal~\cite{Herschkowitz99} show that it
cannot decrease faster than linearly with $\alpha$.  For the Gaussian
scenario, the inequality reads $s_{G}\geq \ln 2 -
(\alpha/2)[B^{2}-A-\ln(1-A)]$.

Before we proceed to study in detail the solution of eq.~\ref{spRG},
we turn to the analysis of the asymptotic behavior of the system.

\subsection{Asymptotics}

The asymptotics of the solution of eq.~\ref{spRG} can be immediately
inferred by carrying out expansions of $F_{B}$ and ${\cal F}$.  In the
vicinity of $R_{G}=0$, if we assume a smooth behavior for
$R_{G}(\alpha)$, the predictions for the Gaussian scenario are:

\begin{eqnarray}
\label{GL:asy:smallalpha1}
B\neq 0 & \Rightarrow & 
R_{G} \simeq \alpha B^{2} \\ 
\label{GL:asy:smallalpha2}
B = 0 & \Rightarrow & 
R_{G} \left\{
\begin{array}{ll}
= 0, & \alpha\leq\alpha_{G} \\
\simeq C_{G} (\alpha-\alpha_{G}), & \alpha\geq\alpha_{G}\; ,
\end{array}
\right.
\end{eqnarray}
where 

\begin{eqnarray}
\label{GL:gauss:asy:constants}
\alpha_{G} & \equiv & \frac{1}{A^{2}} \nonumber \\
C_{G} & \equiv & \frac{A^{2}}{1-A}\; .
\end{eqnarray}
We see that in the so called {\em biased case} $B\neq 0$, it is much
easier to learn.  The {\em unbiased case} $B=0$ presents much more
difficulties for information about vector $\bm{B}$ to be extracted,
due to the intrinsic symmetry $\bm{B}\to -\bm{B}$.  In this case, {\em
retarded learning} occurs~\cite{Biehl94b,Watkin94a}, meaning that a
non-zero macroscopic overlap $R_{G}$ will be obtained only after a
critical number of examples $\alpha_{G}N$ is presented.  For
$\alpha\leq\alpha_{G}$, the entropy saturates its linear bound
exactly~\cite{Herschkowitz99}.
%, leading to the interpretation that the system extracts
%maximal information from the data but is nonetheless unable to produce
%a macroscopic alignment with the preferential
%direction~\cite{Herschkowitz99}.  
This second order phase transition is identical to the one obtained in
the spherical case~\cite{Reimann96b,VandenBroeck96}, revealing that
the binary nature of the preferential direction plays no role in the
poor performance regime.

In the limit $\alpha\to\infty$, on the other hand, the differences 
with respect to the spherical case become pronounced: $R_{G}$ 
approaches 1 exponentially, 

\begin{equation}
\label{GL:asy:largealpha}
1-R_{G}\stackrel{\alpha\to\infty}{\simeq}
\sqrt{\frac{\pi(1-A)}{2\alpha(B^{2}(1-A)+A^{2})}}\,
%\exp\left[
e^{
-\alpha\left(
B^{2}+\frac{A^{2}}{1-A}
\right)/2
}\; ,
%\right]\; .
\end{equation}
as opposed to the power law observed for the spherical case. 
Eq.~\ref{GL:asy:largealpha} also implies an exponential decay to the
entropy, $s_{G}\simeq \alpha(1-R_{G})(B^{2}(1-A)+A^{2})/(2(1-A))$.

These qualitative asymptotic results can be shown to hold for general
distributions ${\cal P}(b)$~\cite{Copelli99b}.  In the following, we
explore the Gaussian scenario in more detail, studying the
behavior of $R_{G}(\alpha)$  away from the asymptotic
regimes.

\subsection{The biased case}
\label{biased}

\begin{figure}[htb!]
\begin{center}
\includegraphics{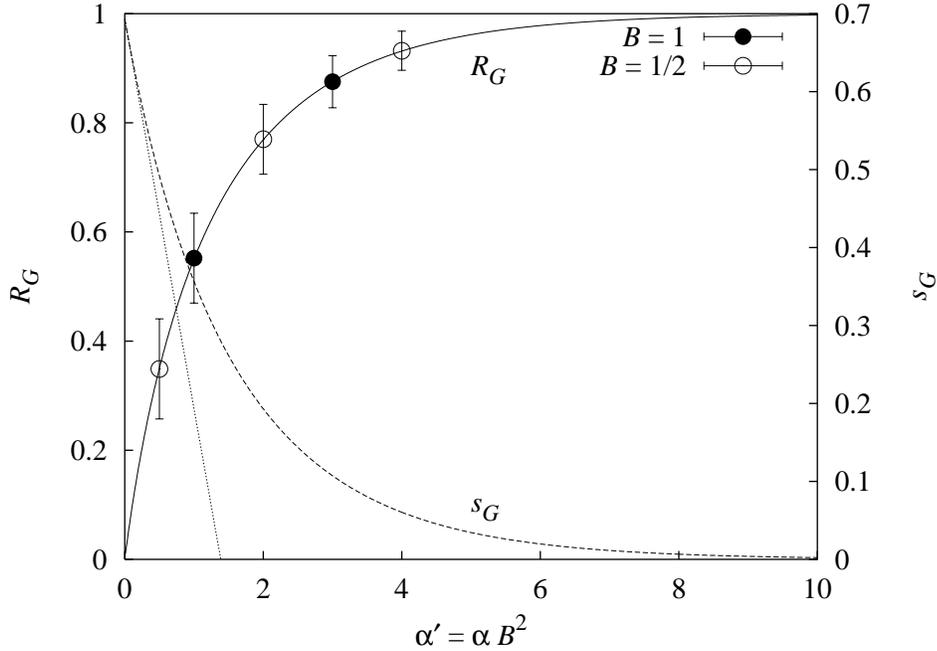}
\caption[short]{Overlap $R_{G}$ (left axis) as a function of
$\alpha^{\prime}$ for $A=0$ (see eq.~\ref{GL:gauss:bias:alphaprime}):
theory (solid line) and simulations with $N=100$ (symbols; error bars
represent one standard deviation, see text for details).  The dashed
line represents the entropy (right axis) while the dotted line shows
the linear bound (right axis).  }
\label{figrrhat}
\end{center}
\end{figure}

The first case to be addressed is $A=0$ with $B\neq 0$.  The non-zero
bias makes sure learning starts off as soon as $\alpha\geq 0$, while
$A=0$ eliminates the dependence of $\hat{R}_{G}$ on $R_{G}$ (see
eqs.~\ref{GL:gauss:sp:sp} and~\ref{spRGhat}), much simplifying the
saddle point equations, which can be solved exactly.  The behavior of
$R_{G}$ is seen to be simply determined by the rescaled variable

\begin{equation}
\label{GL:gauss:bias:alphaprime}
\alpha^{\prime} \equiv \alpha B^{2}\; ,
\end{equation}
namely $R_{G} = F_{B}^{2}\left(\sqrt{\alpha^{\prime}}\right)$.  This
function can be seen in fig.~\ref{figrrhat}.  It shows a linear
increase for small $\alpha^{\prime}$ and an exponential behavior for
$\alpha^{\prime}\to\infty$.  The entropy saturates its linear bound
only in the limit $\alpha^{\prime}\to 0$, approaching zero
exponentially when $\alpha^{\prime}\to\infty$ but remaining otherwise
strictly positive.

%\begin{widetext}

\begin{figure}[hbt!]
\begin{center}
\includegraphics{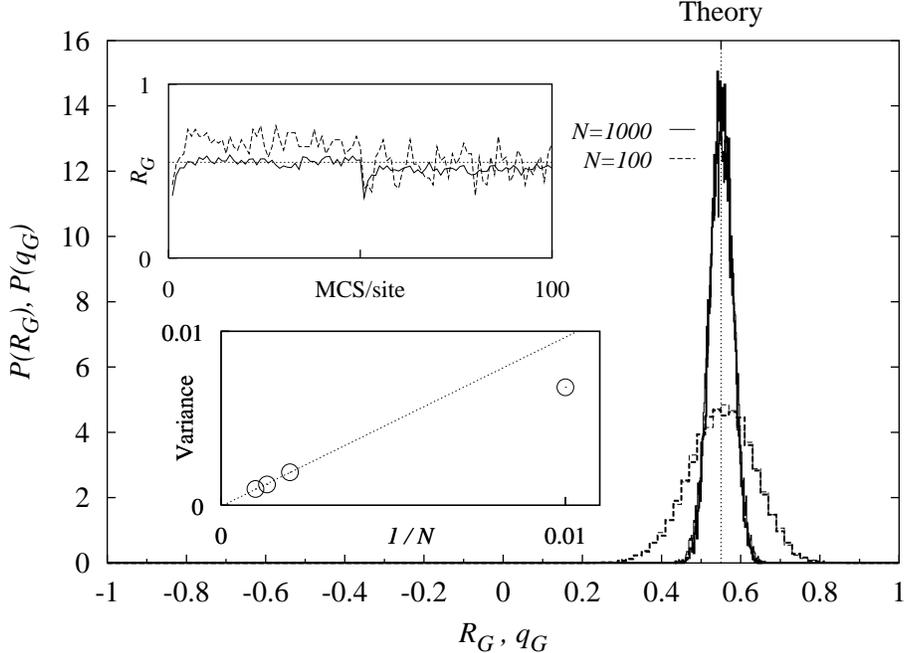}
\caption[short]{Gaussian scenario with $A=0$, $B=1$ and $\alpha=1$. 
Histograms of $q_{G}$ (thick lines) and $R_{G}$
(thin lines) for
$N=100$ (dashed) and $N=1000$ (solid); the vertical line is the
theoretical prediction.  The upper inset shows the Metropolis
$R_{G}$-dynamics for two pattern sets (same legend, see text for
details).  The lower inset presents the variance of the distribution
for $R_{G}$ (symbols) as a function of $1/N$ for $N=100, 500, 750$ and
$1000$; the dotted line is a linear fit of the three leftmost points.}
 \label{figfancyplot}
\end{center}
\end{figure}
%\end{widetext}

Note that $A=0$ means that $b$ has unit variance.  The patterns can
thus be pictured as being distributed in an $N$-dimensional
spherically symmetric cloud, whose displacement $B$ from the origin
conveys the information about $\bm{B}$.

\paragraph*{Simulations}

Binary disordered systems are known to be very hard to simulate due to
the existence of very many local minima.  A noisy dynamics with unity
temperature and general cost function $U$ will typically get stuck in
one of these minima, preventing a proper sampling of the posterior
distribution~\ref{gibbslearning} in an acceptable time.  The Gaussian
scenario with $A=0$ provides an exception to this rule, allowing Gibbs
learning to be very easily implemented with a simple Metropolis
algorithm~\cite{Binder88}.  Since $A=0$ implies a linear function
$U(\lambda)$, the changes in energy can be very quickly calculated
because it depends only on $\bm{J}\cdot\sum_{\mu}\bm{\xi^{\mu}}$.

Fig.~\ref{figrrhat} shows the results for simulations with $N=100$
(the smallest system size simulated) and two values of $B$, checking
the relevance of the variable $\alpha^{\prime}$.  For each pattern set
$D$, 10 samples of $R_{G}$ and $q_{G}$ were measured, after a random
initialization of the system and a warming up of the dynamics (see
further details below).  The whole procedure was repeated for 1000
pattern sets and the standard deviation was calculated over these
10000 samples.

The measurement of $q_{G}$ during simulations is another tool to check
both the property $q_{G}=R_{G}$ and the correctness of the RS {\it
ansatz\/}.  Fig.~\ref{figfancyplot} focus on the second simulated
point of fig.~\ref{figrrhat} ($\alpha^{\prime} = 1$).  It shows
histograms for both $R_{G}$ and $q_{G}$ (measured between pairs of
consecutive samples) which are virtually indistinguishable on the
scale of the figure, with a mean value in excellent agreement with the
theoretical prediction.  The upper inset gives a glimpse of the
Metropolis dynamics: the system is initialized randomly at $t=0$ and
evolves up to $t = 50$ Monte Carlo steps per site (MCS/site), at which
moment a different pattern set is drawn.  The system reaches thermal
equilibrium after ${\cal O}(10)$ MCS/site, which motivated the choice
of safely waiting 100 MCS/site during the simulations before any
measurement was made.  The system was reinitialized after every
measurement of the overlaps.  Note that some pattern sets yield
time-averaged values of $R_{G}$ which deviate from theory (notably the
first one for $N=100$ and the second one for $N=1000$) and only a
second average over the pattern sets gives the right results.  This
reflects the property of self-averaging, which only holds in the
thermodynamic limit (note that deviations from theory are smaller for
larger $N$).  The lower inset shows the typical scaling with
$1/\sqrt{N}$ of the width of the distribution of overlaps.

\subsection{The unbiased case}

\begin{figure}[htb!]
\begin{center}
\includegraphics{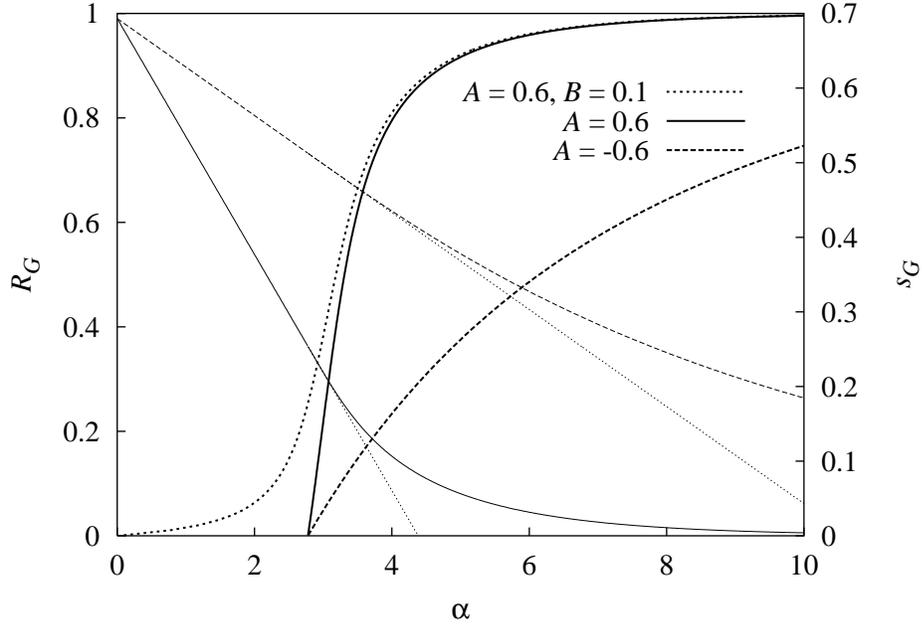}
\caption[short]{$R_{G}$ (thick lines, left axis) and
$s_{G}$ (thin lines, right axis) as functions of 
$\alpha$ for $A=\pm 0.6$ and $B=0$. 
The thin dotted lines correspond to the linear bound, which is exactly
saturated up to the second order phase transition.  A small bias
$B=0.1$ (with $A=0.6$) breaks the symmetry and destroys the second
order phase transition (thick dotted line, left axis, entropy not
shown).}
\label{figrhoalphaA06}
\end{center}
\end{figure}

\begin{figure}[t!]
\begin{center}
\includegraphics{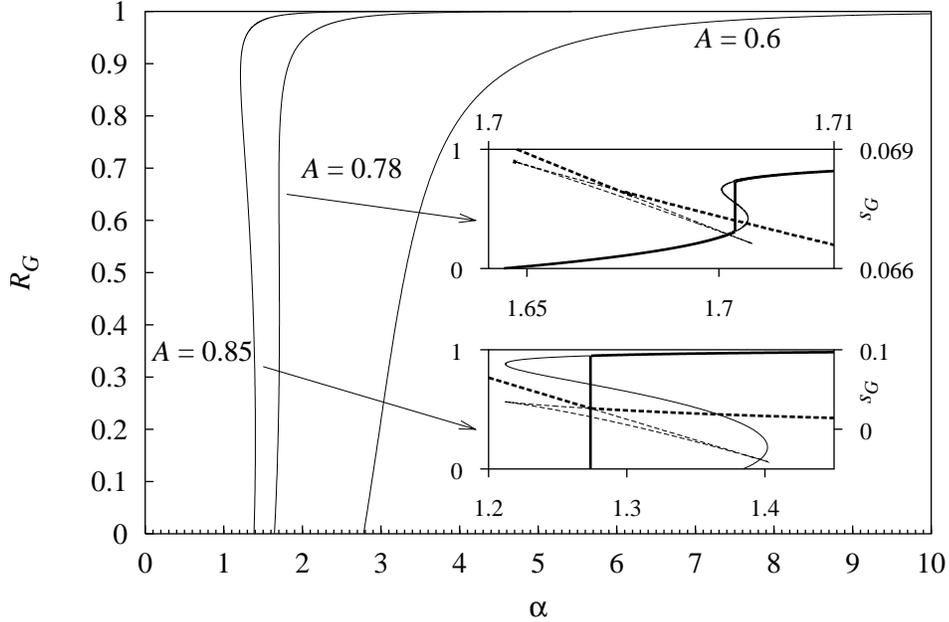}
\caption[short]{Solutions $R_{G}$ of the saddle point
equations~\ref{spRG} and~\ref{GL:gauss:sp:sp} as a function of
$\alpha$ for $B=0$ and three values of $A$.  $s_{G}$ is plotted with
dashed lines and the thermodynamically stable solutions are plotted
with thick lines.  $A=0.78$ (upper inset): $R_{G}$ (left axis) vs. 
$\alpha$ (bottom axis) and $s_{G}$ (right axis) vs.  $\alpha$ (top
axis --- note the different $\alpha$ scale, which zooms in the first
order phase transition); $A=0.85$ (lower inset): $R_{G}$ (left axis)
and $s_{G}$ (right axis) vs.  $\alpha$.}
\label{figbequalzero}
\end{center}
\end{figure}

\begin{figure}[htb!]
\begin{center}
\includegraphics{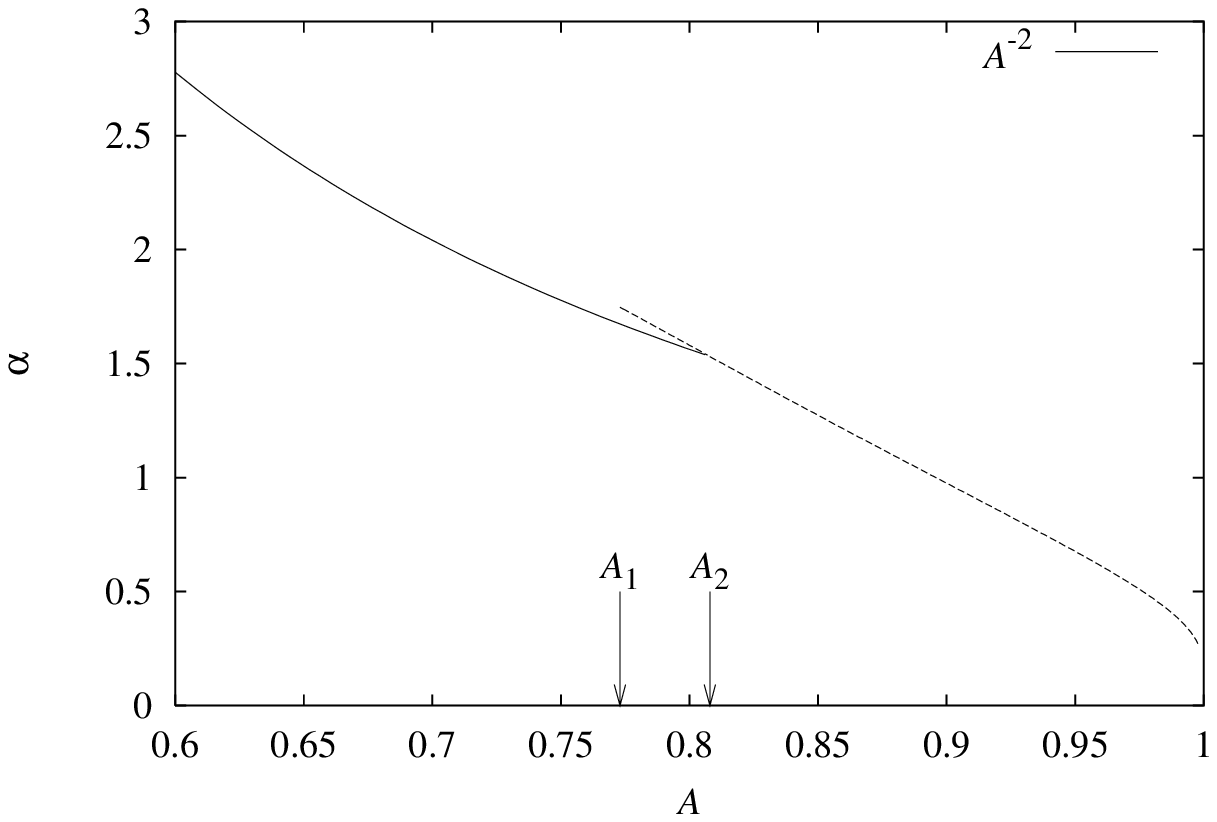}
\caption[short]{Phase diagram for the unbiased case ($B=0$).  Second
order phase transitions occur at $\alpha = \alpha_{G}=A^{-2}$ (solid
line), while first order phase transitions take place at $\alpha =
\alpha_{G}^{(f)}$ (dashed line).  See text for details.  }
 \label{figphasedia}
\end{center}
\end{figure}

When $B=0$, retarded learning is expected to occur, according to
eq.~\ref{GL:asy:smallalpha2}.  Fig.~\ref{figrhoalphaA06} shows the
solution of the $R_{G}$ saddle point equation for two values of $A$,
namely 0.6 and $-0.6$.  In both cases, a second order phase transition
occurs at the critical value $\alpha_{G}$ predicted by
eq.~\ref{GL:gauss:asy:constants} and the entropy saturates exactly the
linear bound before the phase transition.  Based on the relation
between $s_{G}$ and $i$~\cite{Herschkowitz99}, the retarded learning
phase transition can be interpreted as follows: for
$\alpha\leq\alpha_{G}$, the system extracts maximal information from
each pattern but is nonetheless unable to obtain a non-zero alignment
$R_{G}$ with the preferential direction $\bm{B}$.  Only at
$\alpha=\alpha_{G}$ does $R_{G}$ depart from zero, which on its turn
immediately gives an increasing degree of redundancy (measured by the
deviation of $s_{G}$ from its linear bound) to the patterns coming
thereafter.  Fig.~\ref{figrhoalphaA06} also shows the effect of a
small bias $B=0.1$ in an otherwise symmetric distribution: for
sufficiently large $\alpha$ (say, $\alpha\gg\alpha_{G}$), the effect
is negligible, but for small $\alpha$ the broken symmetry destroys the
second order phase transition.

It is interesting to note in fig.~\ref{figrhoalphaA06} that even
though the phase transition for $A=-0.6$ and $A=0.6$ occurs at the
same critical value, the overlap increases much slower in the former
case than in the latter.  Recalling the definition of $A$
(eqs.~\ref{pofb}-\ref{UGauss}), this means that prolate Gaussian
distributions ($N$-dimensional ``cigars''~\cite{Biehl94b}) convey less
information about the preferential direction than oblate distributions
($N$-dimensional ``pancakes''~\cite{Biehl94b}) for the same absolute
value of $A$.

However, the second order phase transition at $\alpha_{G} = A^{-2}$ is
not the only interesting phenomenon for this model.  First order phase
transitions are also possible, depending on the value of $A$.  They
can occur in two situations: either for $\alpha>\alpha_{G}$, in which
case two consecutive phase transitions take place during learning (a
second-order one followed by a first-order one), or
$\alpha<\alpha_{G}$, in which case the asymptotic result
eq.~\ref{GL:asy:smallalpha2} is overridden.  The first order phase
transition appears when there is more than one solution to the saddle
point equation.  In such cases the solution with minimal free energy
has maximal probability of occurrence, being thus the
thermodynamically stable state.  Such first order phase transitions
have been found for the spherical case with a two-peaked
distribution~\cite{Buhot98}, but not with the Gaussian
scenario~\cite{Reimann96b}, which shows that they are due to the
discrete nature of the search space in this case.

An overview of this phenomenology is presented in
fig.~\ref{figbequalzero}.  It shows the three typical behaviors that
occur for $B=0$.  For comparison, the case $A=0.6$ plotted in
fig.~\ref{figbequalzero} is shown again, as an example of a parameter
region where there is only a second order phase transition (at
$\alpha_{G} = 2.78$).  For $A=0.78$ the second order phase transition
at $\alpha_{G}=1.643$ is followed by a first order phase transition at
$\alpha_{G}^{(f)} =1.704$ (upper inset, lower axis), while for
$A=0.85$ only a first order phase transition takes place at $\alpha =
1.27$, overriding the second order phase transition at $\alpha = 1.38$
(lower inset) which was predicted on asymptotics and smoothness
grounds.  Note that none of these first order phase transitions can be
predicted by the asymptotic expansion, eq.~\ref{GL:asy:smallalpha2}. 
It is also interesting to observe that some solutions of the saddle
point equation may violate the linear bound and/or the positivity of
the entropy (notably $A=0.85$ in fig.~\ref{figbequalzero}).  However,
it turns out that these branches are always thermodynamically
unstable, while the stable solutions satisfy all the requirements.

The whole phase diagram for $B=0$ is shown in fig.~\ref{figphasedia}. 
For $A>A_{1} \simeq 0.773$, a first order phase transition takes place
at the line $\alpha_{G}^{(f)}(A)$, after the second order one has
already occurred.  For increasing $A$, $\alpha_{G}^{(f)}(A)$ gets
closer and closer to $\alpha_{G}(A)$, until there is finally a
collapse at $A=A_{2}\simeq 0.808$.  For larger values of $A$, only the
first order phase transition occurs.

\section{Optimal learning: the Bayesian perspective}
\label{Bayes}

We now switch to the following question: given the $\alpha N$ data
vectors and the prior information about $\bm{B}$, what is the {\em
best\/} performance $R$ one could possibly attain with a vector
$\bm{J}$?  Watkin and Nadal~\cite{Watkin94a} answered this question in
a Bayesian framework by defining {\em optimal learning\/} (see
also~\cite{Opper91a,Watkin93b}).  We briefly review their reasoning
here and extend it to take into account the binary nature of the
vectors.

We define the quality measure ${\cal Q}(\bm{B},\bm{J})\equiv
\bm{B}\cdot\bm{J}/N$, which quantifies how well $\bm{B}$ is approximated
by any candidate vector $\bm{J}$ satisfying $\bm{J}\cdot\bm{J}=N$. 
Since $\bm{B}$ is unknown, ${\cal Q}$ is formally inaccessible.  But
one can take its average with respect to the posterior
distribution~\ref{posterior}, leading to $\tilde{\cal Q}(\bm{J},D)
\equiv \int d\bm{B}\, {\cal Q}(\bm{B},\bm{J}) p(\bm{B}|D)$. 
$\tilde{\cal Q}$ is then a formally accessible {\it bona fide\/}
quantity which can be used to measure the performance of $\bm{J}$.

Optimal learning is {\em defined\/} as constructing a vector
$\bm{J}_{B}$ which maximizes $\tilde{\cal Q}$.  The linearity of
${\cal Q}$ in $\bm{J}$ immediately implies

\begin{equation}
\label{Qtilde}
\tilde{\cal Q}(\bm{J},D) = N^{-1}\bm{J}\cdot\int d\bm{B}\, \bm{B}\, 
p(\bm{B}|D)\; ,
\end{equation}
leading on its turn to

\begin{equation}
\label{JB}
\bm{J}_{B} = \frac{1}{\sqrt{R_{G}}}\int d\bm{B}\, \bm{B}\, p(\bm{B}|D)\; ,
\end{equation}
where the $\sqrt{R_{G}}$ factor guarantees the proper normalization of
$\bm{J}_{B}$.  This is the so-called Bayesian vector, which is the
center of mass of the Gibbs ensemble.  In the thermodynamic limit, its
performance $R_{B}\equiv \bm{B}\cdot\bm{J}_{B}/N$ is shown to be
simply related to that of Gibbs learning~\cite{Watkin93b,Watkin94a}:
$R_{B}=\sqrt{R_{G}}$.

\subsection{The best binary}

Up to now the reasoning is fairly general.  The whole procedure can
actually be carried out without explicitly mentioning what the prior
distribution $P(\bm{B})$ is.  For clarity, in the following
$\bm{J}_{B}$ will specifically denote the Bayesian vector for a {\em
binary\/} prior.  Note, however, that despite being the center of mass
of the ensemble of {\em binary\/} vectors sampled from the posterior
distribution, $\bm{J}_{B}$ has real-valued
components~\cite{Copelli99b}, in general.

One would therefore like to address the next question: what is the
{\em best binary\/} vector one can construct? In other words,
what is the binary vector --- inferable from the data --- that
outperforms --- on average --- any other binary vector in
approximating $\bm{B}$?  The answer is again
straightforward~\cite{Watkin93a}: the vector $\bm{J}_{bb}$ which
maximizes $\tilde{\cal Q}$ {\em among the binary vectors} is simply
obtained by the clipping prescription, namely $[\bm{J}_{bb}]_{j} =
\mbox{sign}([\bm{J}_{B}]_{j})$, $j=1,\ldots,N$ or, in shorthand
notation,

\begin{equation}
\label{Jbb}
\bm{J}_{bb} = \mbox{clip}(\bm{J}_{B})\; .
\end{equation}
This can be easily checked by noting that the quantity to be maximized
(the r.h.s. of eq.~\ref{Qtilde}) is proportional to
$\sum_{j=1}^{N}J_{j}[\bm{J}_{B}]_{j}$.  In what follows, $\bm{J}_{bb}$
is called the {\it best binary\/} vector.

Summarizing, if $\bm{B}$ is known to be binary, $\bm{J}_{B}$ is the
best estimator one can provide.  But if the estimator is required to
be binary as well, then $\bm{J}_{bb}$ is the optimal choice.  

The proof that $\bm{J}_{B}$ and $\bm{J}_{bb}$ are
optimal estimators in their
respective spaces, is relatively simple~\cite{Reimann96a}. What we
show below is that maximal $\tilde{\cal Q}$ implies maximal
alignment $R$ with $\bm{B}$, in the
thermodynamic limit.
For the best binary, one departs from the inequality $\tilde{\cal
Q}(\bm{J}_{bb})-\tilde{\cal Q}(\bm{J})\geq 0$,
$\forall\bm{J}\in\{-1,+1\}^{N}$ and takes the average with respect to
the data distribution:

\begin{eqnarray}
\label{demonstration}
& &\int dD\, P(D)\left[ \tilde{\cal Q}(\bm{J}_{bb}) - \tilde{\cal 
Q}(\bm{J}) \right] = \nonumber \\
& = & \int d\bm{B}\,P_{b}(\bm{B}) \int dD\, P(D|\bm{B})
\left[\frac{\bm{B}\cdot\bm{J}_{bb}}{N} -
\frac{\bm{B}\cdot\bm{J}}{N}\right] \geq 0\; .
\end{eqnarray}
If we now assume that $R_{bb}\equiv \bm{B}\cdot\bm{J}_{bb}/N$ 
and $R= \bm{B}\cdot\bm{J}_{bb}/N$ are self-averaging, the average
over the data can be bypassed and one obtains

\begin{equation}
\label{demonstration2}
\int d\bm{B}\,P_{b}(\bm{B})\left[ R_{bb} - R \right]
\stackrel{N\to\infty}{\geq} 0\; ,
\end{equation}
Finally, one notices that
$P_{b}(\bm{B})$ is a uniform prior, making no distinction between any
particular binary vector (this is reflected, for instance, in the free
energy~\ref{freebin} being independent of the particular choice of
$\bm{B}$): this allows the last average in eq.~\ref{demonstration2} to
be bypassed as well, leading to the stronger upper bound $R_{bb}\geq
R$.

\subsection{Performance and Simulations}
\label{persim}

The performance $R_{bb}$ of the best binary vector can be explicitly
calculated by extending previously obtained results for the clipping
prescription~\cite{Bolle95a,Schietse95}.  In~\cite{Schietse95},
Schietse {\it et al.\/} study the effect of a general transformation
$\tilde{J}_{j} = \sqrt{N}\phi(J_{j})/\sqrt{\sum_{i}\phi^{2}(J_{i})}$
on the components of a properly normalized continuous vector $\bm{J}$
satisfying $\bm{B}\cdot\bm{J}/N=R$.  If $\bm{B}$ is binary, $\phi$ is
odd and $\tilde{R}\equiv\bm{B}\cdot\tilde{\bm{J}}/N$ is
self-averaging, then the following relation follows:

\begin{equation}
\label{rtilde}
\tilde{R} = \frac{\int P(x)\,\phi(x)\,dx}
{\left[\int P(x)\,\phi^{2}(x)\,dx
\right]^{1/2}}\; ,
\end{equation}
where the variable $x\equiv B_{1}J_{1}$ is expected to be distributed
independently of the index, because of the permutation symmetry among
the axes.

\begin{figure}[htb!]
\begin{center}
\includegraphics{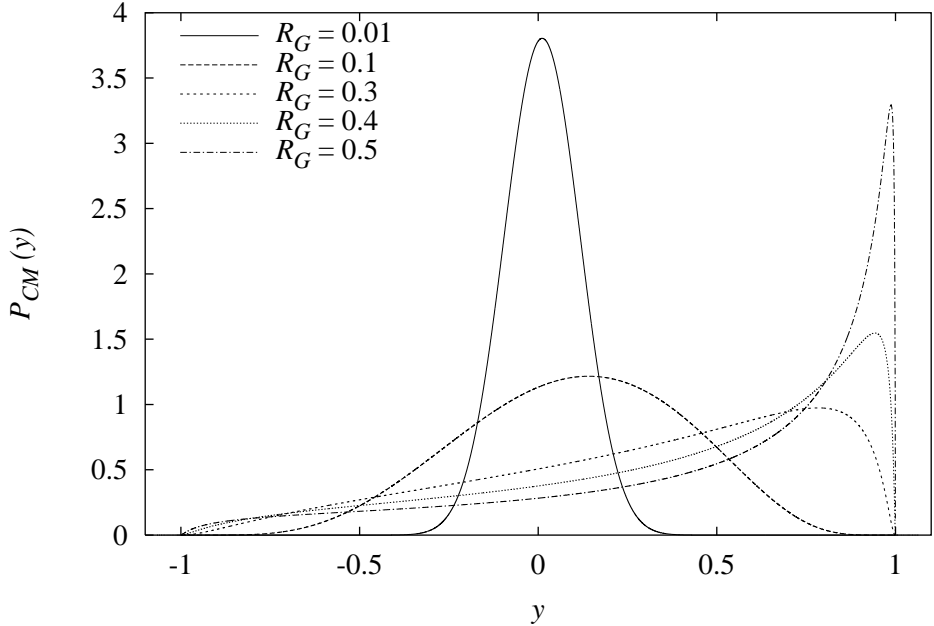}
\caption[short]{Probability distribution of $y =
B_{1}\left<J_{1}\right>_{\bm{J}}$ for different values of $R_{G}$.}
\label{fignewpy}
\end{center}
\end{figure}

We are left then with the problem of calculating $P(x)$, after which
eq.~\ref{rtilde} can be applied for $\phi(x)=\mbox{sign}(x)$,
providing $R_{bb}$ as a function of $R_{G}$.  If $\bm{J}$ is
uniformly distributed on the cone $\bm{B}\cdot\bm{J}=NR$, then $P(x)$
is just a Gaussian with mean $R$ and variance
$1-R^{2}$~\cite{Schietse95}.  However this can hardly be expected to
hold for the Bayesian vector, since it is a sum of Ising vectors.  One
would naively expect $\bm{J}_{B}$ to be closer to the corners of the
$N$-hypercube instead.  To obtain the relevant $P_{CM}(x)\equiv
P(x=B_{1}[\bm{J}_{B}]_{1})$, we calculate the $m$-th quenched moment
of $[\bm{J}_{B}]_{1}$,

\begin{equation}
\label{mthmoment}
\left< ([\bm{J}_{B}]_{1})^{m}\right>_{D} = 
\frac{1}{R_{G}^{m/2}}\left< 
Z^{-m}\left(
\int d\bm{J}\,P_{b}(\bm{J})\, e^{-\beta\sum_{\mu}^{\alpha 
N}U(\lambda_{\mu})}J_{1}
\right)^{m}
\right>_{D}\; .
\end{equation}
The above expression can be evaluated again with the use of the 
replica trick. The replica symmetric result is 

\begin{equation}
\label{mthmoment:final1}
\left< ([\bm{J}_{B}]_{1})^{m}\right>_{D} = \frac{1}{(R_{G}(\alpha))^{m/2}}\int
{\cal D}z\, \left[ \tanh\left(
z\sqrt{\hat{R}_{G}(\alpha)}+\hat{R}_{G}(\alpha)B_{1} \right) \right]^{m}\; ,
\end{equation}
or, equivalently, 

\begin{equation}
\label{mthmoment:final2}
\left<x^{m}\right>_{D} = 
\frac{1}{(R_{G}(\alpha))^{m/2}}
\int {\cal D}z\, 
\left[ \tanh\left( z\sqrt{\hat{R}_{G}(\alpha)}+\hat{R}_{G}(\alpha) \right) 
\right]^{m}\; ,
\end{equation}
where the values of ${R}_{G}(\alpha)$ and $\hat{R}_{G}(\alpha)$ should
be taken at the solution of the saddle point equations~\ref{spRG}
and~\ref{spRGhat} for Gibbs learning.  Notice that the passage from
eq.~\ref{mthmoment:final1} to eq.~\ref{mthmoment:final2} is valid {\em
only\/} if $\bm{B}$ is binary.  From eq.~\ref{mthmoment:final2} one
immediately rewrites the probability distribution $P_{CM}(x)$, by
identifying a change of stochastic variables $x =
R_{G}^{-1/2}\tanh\left(z\sqrt{\hat{R}_{G}}+\hat{R}_{G}\right)$, with
$z$ normally distributed:

\begin{equation}
\label{px}
P_{CM}(x)  =  \frac{\sqrt{R_{G}}}{\sqrt{2\pi\hat{R}_{G}}(1-R_{G}x^{2})}
\,\exp\frac{-1}{2\hat{R}_{G}}
\left[
\frac{1}{2}\ln\left( 
               \frac{1+\sqrt{R_{G}}x}{1-\sqrt{R_{G}}x}
               \right)
- \hat{R}_{G}
\right]^{2}\; .
\end{equation}
Note that, since $R_{G}$ and $\hat{R}_{G}$ are simply related to each
other (eqs.~\ref{spRG} and~\ref{spRGhat}), $P_{CM}(x)$ can always be
parametrized in function of $R_{G}$ only.  In fig.~\ref{fignewpy}, the
probability distribution of $y\equiv x\sqrt{R_{G}}$ is plotted for
different values of $R_{G}$, illustrating the fact that $|y|\leq 1$.

\begin{figure}[h!bt]
\begin{center}
\includegraphics{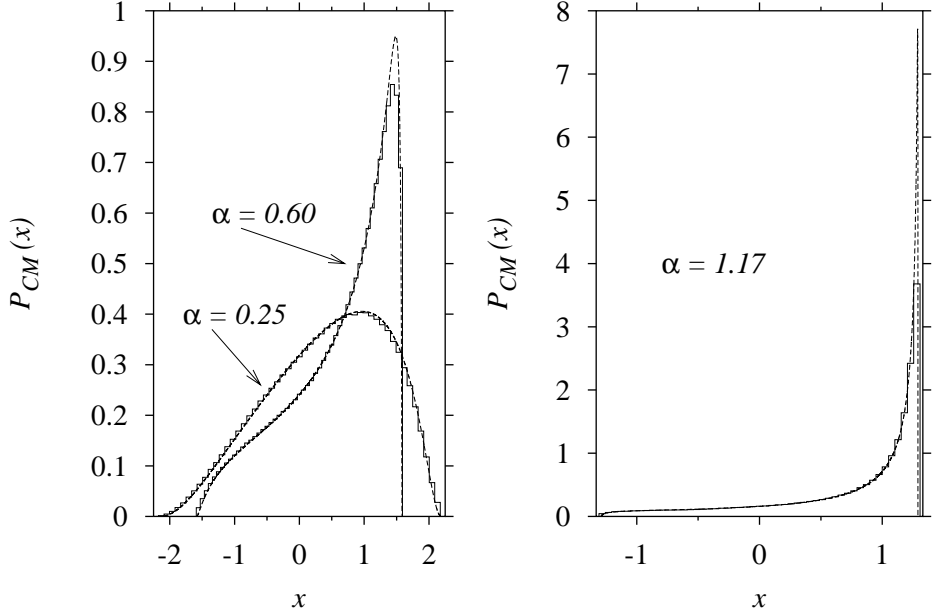}
\caption[short]{Gaussian scenario with $A=0$ and $B=1$.  Probability
distribution of $x = B_{1}[\bm{J}_{B}]_{1}$ for $\alpha = 0.25$,
$\alpha = 0.6$ (left plot) and $\alpha = 1.17$ (right plot) or,
correspondingly, $R_{G} = 0.2$, $R_{G} = 0.4$ and $R_{G} = 0.6$. 
Solid line: simulations.  Dashed line: theory (eq.~\ref{px}).}
\label{figsimpx}
\end{center}
\end{figure}

Eq.~\ref{px} should be compared to the Gaussian distribution obtained
in~\cite{Schietse95}.  It shows that $\bm{J}_{B}$ is indeed closer to
the corners of the $N$-hypercube, optimally incorporating the
information that $\bm{B}$ is binary.

We have run simulations for $A=0$ and $B=1$ as described in
section~\ref{biased}.  For a system size $N=500$, the center of mass
was constructed with $n=50$ samplers, being normalized afterwards. 
Each component of $\bm{B}$ and $\bm{J}_{B}$ was used to measure $x$,
the procedure being repeated 100 times for each of the 100 pattern
sets.  A comparison between the resulting histogram and the
theoretical prediction can be seen in fig.~\ref{figsimpx}.  The good
agreement shows that eq.~\ref{px} correctly describes the statistical
properties of the Bayesian vector.

We can finally proceed to calculate the performance $R_{bb}$ of the
best binary vector. Making use of eqs.~\ref{rtilde} and~\ref{px}, we 
make a change of variables to obtain 

\begin{eqnarray}
\label{JBclip1}
R_{bb} & = & \int P_{CM}(x)\,\mbox{sign}(x)\,dx %\nonumber \\
% & = & \int{\cal D}z\,\mbox{sign}
%\left( z\sqrt{\hat{R}_{G}(\alpha)}+\hat{R}_{G}(\alpha) \right) \nonumber \\
% & = & 
=1 - 2H\left(\sqrt{\hat{R}_{G}(\alpha)} 
 \right)\; ,
\end{eqnarray}
where $H(x) \equiv \int_{x}^{\infty}{\cal D}z$. Making use of the 
relation between $R_{G}$ and $\hat{R}_{G}$, we finally write 

\begin{eqnarray}
\label{JBclip}
R_{bb} 
 & = & 1-2H\left(   F_{B}^{-1}(R_{B}) \right) = 1-2H\left( 
 F_{B}^{-1}\left( \sqrt{R_{G}}\right) \right) \\
\label{Rbbpotential}
 & = & 1-2H\left( {\cal F}\left(R_{B}\right)\right)
   =  1-2H\left( \sqrt{\hat{R}_{G}} \right)\; ,
\end{eqnarray}
where $F_{B}^{-1}$ is the inverse of $F_{B}$. 

\begin{figure}[tbh!]
\begin{center}
\includegraphics{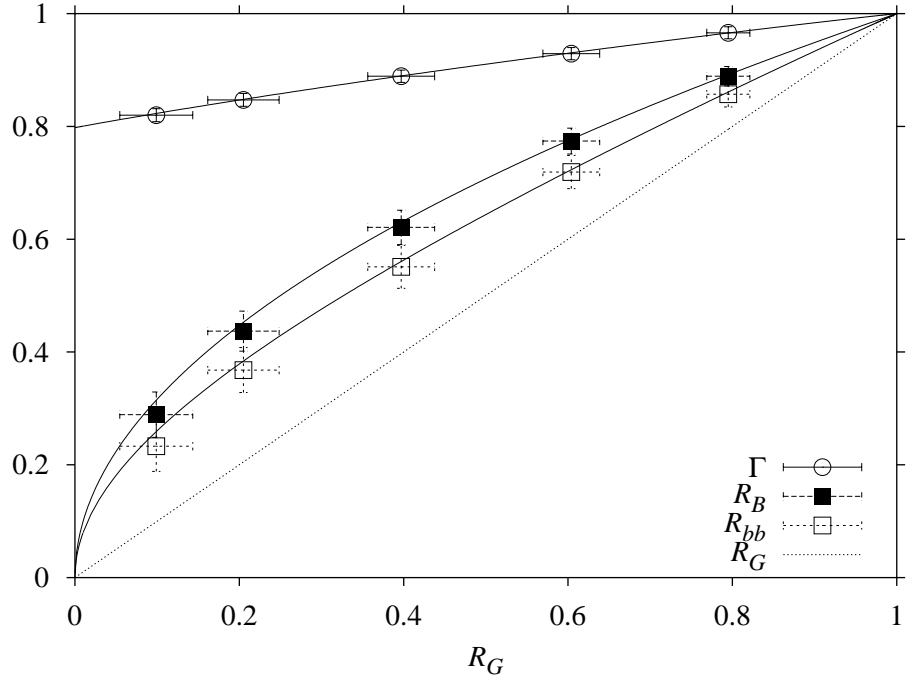}
\caption[short]{\label{figoverlapssim}$R_{bb}$, $R_{B}$ and $\Gamma$:
comparison between simulations and theory, for different values of
$\alpha$.  Error bars represent one standard deviation.  The diagonal
is plotted for comparison.}
\end{center}
\end{figure}

\begin{figure}[hbt!]
\begin{center}
\includegraphics{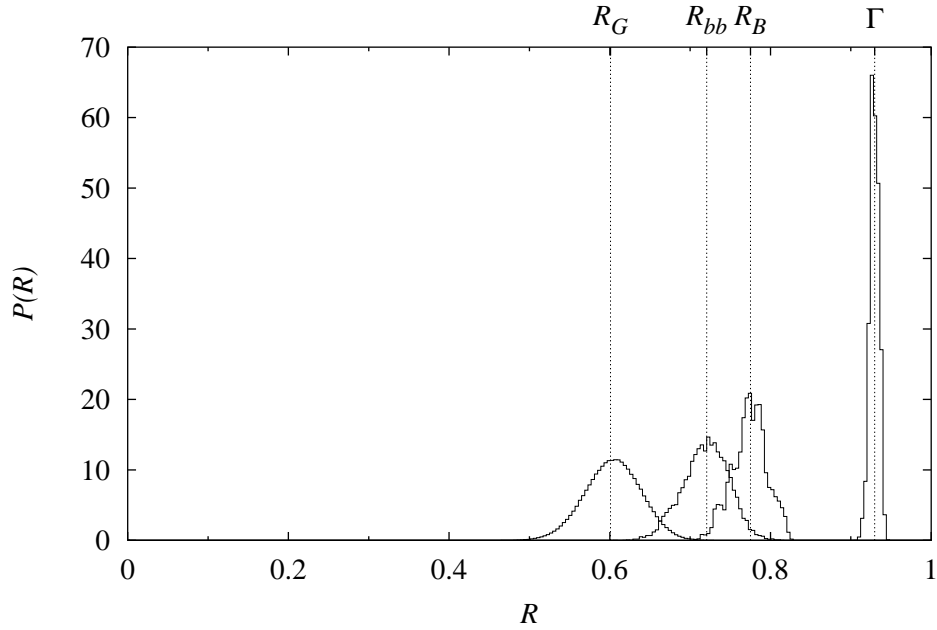}
\caption[short]{\label{fighistoverlapsalpha117}Histograms of
the overlaps $R_{G}$, $R_{B}$, $R_{bb}$ and $\Gamma$ for $\alpha =
1.17$.  The vertical lines show the theoretical predictions.}
\end{center}
\end{figure}

Eq.~\ref{JBclip} expresses an upper bound for binary candidate vectors
$\bm{J}$ in approximating $\bm{B}$, satisfying two obvious
inequalities: $R_{G}\leq R_{bb} \leq R_{B}$.  The asymptotic behavior
of $R_{bb}$ can always be written in terms of $R_{G}$: in the poor
performance regime ($R_{G}\to 0$), one recovers
previous results for clipping a
spherical vector~\cite{Schietse95,VandenBroeck93}, $R_{bb}\simeq
\sqrt{2R_{G}/\pi}$, while in the large $\alpha$
regime ($R_{G}\to 0$) a faster
exponential decay is achieved than with Gibbs learning:
$1-R_{bb}\simeq (2/\pi)(1-R_{G})$~\cite{Copelli99b}.

As a spin-off of the calculation, we have also obtained the overlap
between the Bayesian vector and its clipped counterpart,
$\Gamma\equiv\bm{J}_{B}\cdot\bm{J}_{bb}/N =
N^{-1}\sum_{j}^{N}|[\bm{J}_{B}]_{j}|$.  This quantity can be easily
computed,

\begin{eqnarray}
\label{Gamma}
\Gamma & = & \frac{1}{R_{B}}\int{\cal D}z\, \left|
\tanh\left( z\sqrt{\hat{R}_{G}(\alpha)}+\hat{R}_{G}(\alpha) \right) 
\right|
\nonumber \\ & = &
\frac{R_{bb}}{R_{B}}\; ,
\end{eqnarray}
if one notices the counterintuitive identity $\int{\cal
D}z\,|\tanh(za+a^{2})| = \int{\cal D}z\,\mbox{sign}(za+a^{2})$,
$\forall a$, which is proved in the appendix.  The simple result
$\Gamma=R_{bb}/R_{B}$ immediately implies the equality $\left(
\bm{J}_{bb} - \Gamma\bm{J}_{B} \right)\cdot \left( \bm{J}_{B} -
R_{B}\bm{B} \right) = 0$, for which we still have not found a deeper
interpretation.

The curves $R_{bb}$, $R_{B}$ and $\Gamma$ as functions of $R_{G}$ are
plotted on fig.~\ref{figoverlapssim}, together with results for
simulations with the same parameters as those of fig.~\ref{figsimpx}. 
The data is in excellent agreement with the theoretical results,
errors generally remaining below the margin of one standard deviation. 
Note that $\Gamma\geq\sqrt{2/\pi}$, with equality holding only for
$R_{G}\to 0$.  This result is another confirmation of the picture that
$\bm{J}_{B}$ lies closer to the binary vectors, since $\sqrt{2/\pi}$
is the overlap between a continuous vector isotropically sampled from
the $N$-hypersphere and its clipped counterpart. 
Fig.~\ref{fighistoverlapsalpha117} zooms in the fourth column of
points of fig.~\ref{figoverlapssim} ($\alpha=1.17$), showing the
histograms of the overlaps.  One observes that the distribution of
$\Gamma$ is much sharper than the other ones, while the statistics of
$R_{G}$ is better because it has $n=50$ times more samples.

\section{Transforming the components}
\label{transform}

\subsection{General results}

Since sampling the binary Gibbsian vectors is usually a very difficult
task, the construction of the Bayesian vector according to the center
of mass recipe is not always possible, in practice.  Alternative
methods should therefore be developed for approximating the $R_{B}$
and $R_{bb}$ performances.  One such method is the technique of
transforming the components of a previously obtained {\em spherical\/}
vector, as described in section~\ref{persim}.  In the following, we
first derive general results (for any function $U$) and subsequently
look at the Gaussian scenario in detail.

A natural choice for the vector to be transformed is $\bm{J}_{opt}^s$,
which can be obtained by minimizing --- in the $N$-hypersphere --- an
optimally constructed cost
function~\cite{Reimann96b,VandenBroeck96, Buhot98, Gordon98}
${\cal H}=\sum_{\mu}V_{opt}^{s}(\lambda_{\mu})$.  It attains the
Bayes-optimal performance $R_{opt}^s\equiv
\bm{B}\cdot\bm{J}_{opt}^s/N$ {\em for the spherical case\/}, which
satisfies

\begin{equation}
\label{Ropts}
R_{opt}^s = F_{s}\left({\cal F}(R_{opt}^s)\right)\; ,
\end{equation}
where $F_{s}(x) \equiv x/\sqrt{1+x^{2}}$.  Note that $\bm{J}_{opt}^s$
saturates the performance of the center of mass of the Gibbs ensemble
{\em for a spherical prior\/}.  Eq.~\ref{Ropts} should be compared to
the performance of $\bm{J}_{B}$, which obeys

\begin{equation}
\label{RB}
R_{B} = F_{B}\left({\cal F}(R_{B})\right)\; .
\end{equation}
While $R_{opt}^s\simeq R_{B}$ for small $\alpha$, the differences
between $\bm{J}_{B}$ and $\bm{J}_{opt}^s$ are clearly manifested in
the asymptotic behavior for large $\alpha$, with $R_{opt}^s$
approaching unity with a power law~\cite{Reimann96b,VandenBroeck96}
instead of the exponential decay of eq.~\ref{GL:asy:largealpha}.  

\begin{figure}[th!]
\begin{center}
\includegraphics{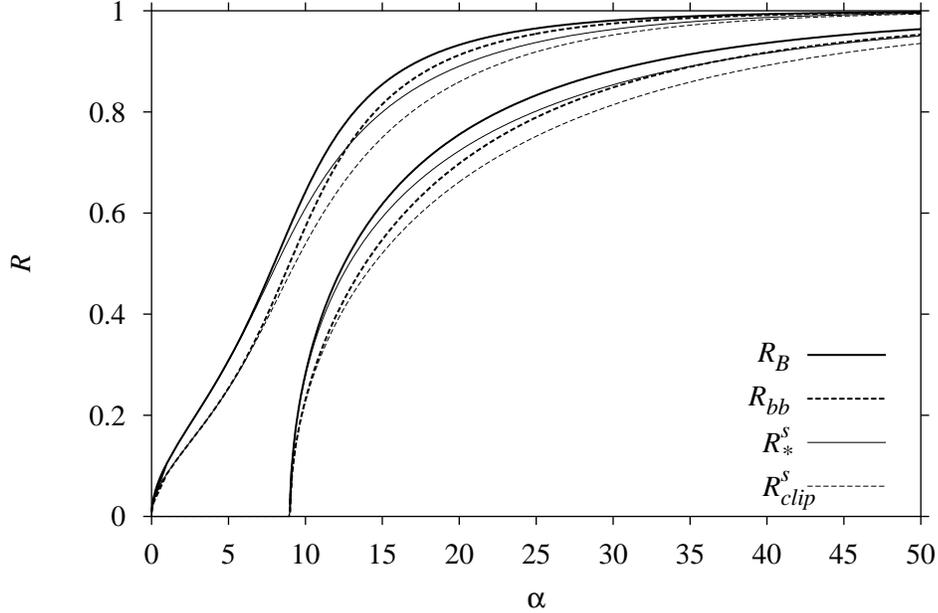}
\caption[short]{Overlaps as functions of $\alpha$ for two choices of
parameters in the Gaussian scenario: $A = 1/3$ with $B = 0.1$ (upper
curves) and $A = -1/3$ with $B = 0$ (lower curves).  The upper bounds
$R_{B}$ (solid) and $R_{bb}$ (dashed) are depicted with thick lines,
while the approximations $R_{*}^{s}$ (solid) and $R_{clip}^{s}$
(dashed) are plotted with thin lines.}
\label{Fig:compareOptTransfSpherical}
\end{center}
\end{figure}

\begin{figure}[h!t]
\begin{center}
\includegraphics{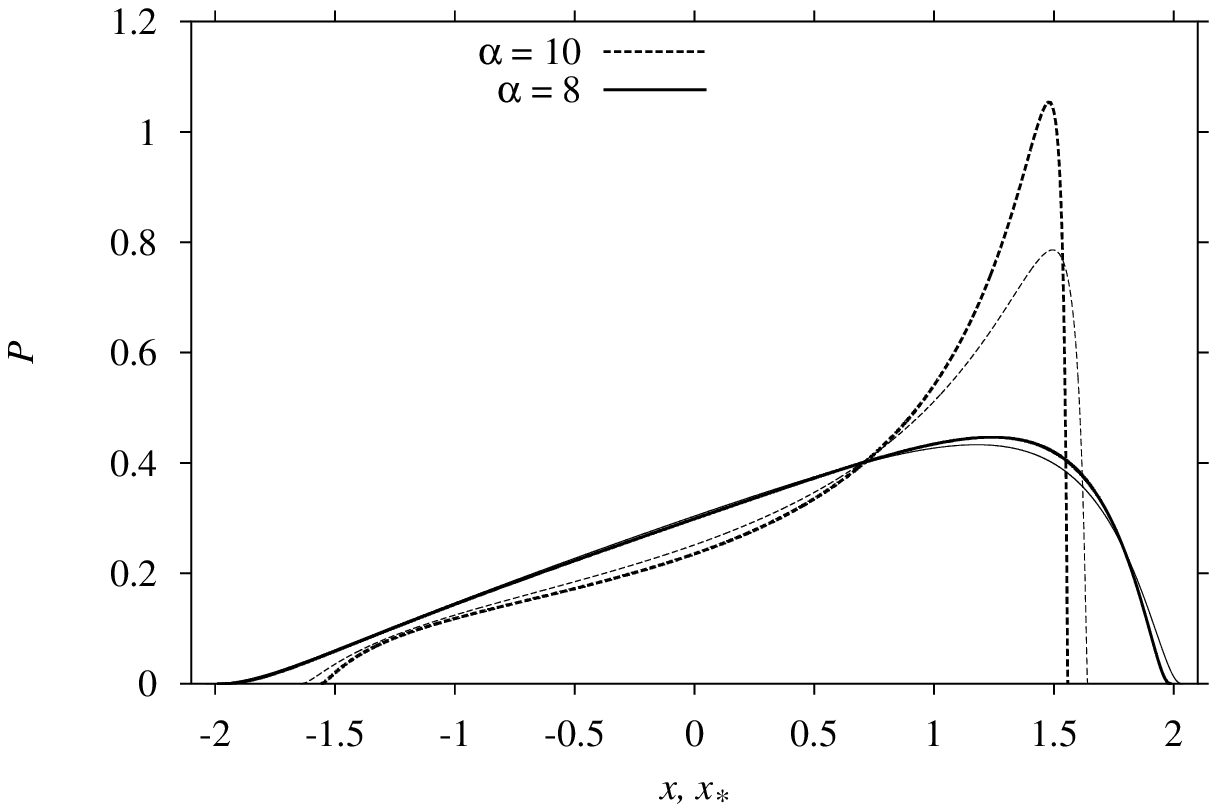}
\caption[short]{Distributions $P_{CM}(x)$ (thick) and $P(x_{*})$
(thin) according to eqs.~\ref{px} and eq.~\ref{pxstar},
respectively.  The values $\alpha = 8$ (solid) and $\alpha=10$
(dashed) refer to the Gaussian scenario with $A=1/3$ and $B=0.1$ (see
fig.~\ref{Fig:compareOptTransfSpherical}).}
\label{Fig:pxvspxprime}
\end{center}
\end{figure}

We would like to depart from $\bm{J}_{opt}^s$ and obtain
approximations to both $\bm{J}_{bb}$ and $\bm{J}_{B}$.  The first one
is obtained by clipping, $\bm{J}_{clip}^{s}\equiv
\mbox{clip}(\bm{J}_{opt}^s)$.  The second one relies on an optimal
transformation~\cite{Schietse95} $\phi^{*}(x)\equiv
(P(x)-P(-x))/(P(x)+P(-x))$, which maximizes the transformed
overlap\footnote{Consistently, the optimal transformation for
$P(x)=P_{CM}(x)$ becomes $\phi^{*}(x)\propto x$, that is, no
improvement is possible.}.  The vector obtained by such a
transformation on $\bm{J}_{opt}^s$ is denoted by $\bm{J}_{*}^{s}$,
thus $[\bm{J}_{*}^{s}]_{j} =
\phi^{*}([\bm{J}_{opt}^s]_{j})/R_{*}^{s}$, $j=1,\ldots,N$, where
$R_{*}^{s}\equiv \bm{B}\cdot\bm{J}_{*}^{s}/N$.

Since $\bm{J}_{opt}^s$ contains no information about the binary nature
of $\bm{B}$, the results of Schietse {\it et al.\/} can be directly
applied to render $R_{clip}^{s}\equiv \bm{B}\cdot\bm{J}_{clip}^{s}/N$
and $R_{*}^{s}$.  In this case, $P(x)$ is Gaussian and one
obtains~\cite{Schietse95}

\begin{eqnarray}
\label{compareRopts}
R_{clip}^{s} & = & 1-2H\left({\cal F}(R_{opt}^{s})\right) \\
\label{compareRstars}
R_{*}^{s} & = & F_{B}\left( {\cal F}(R_{opt}^{s}) \right)\; .
\end{eqnarray}

We would like to compare eqs.~\ref{JBclip} and~\ref{RB} with
eqs.~\ref{compareRopts} and~\ref{compareRstars}, respectively. 
Despite their resemblance in form, one notices that the former should
be solved, while the latter just map the solution of eq.~\ref{Ropts}. 
In order to compare the equations, one should first note that
$F_{B}(x)\geq F_{s}(x)$, $\forall x\geq 0$.  Since $\partial{\cal
F}/\partial R\geq 0$, in general $R_{B}\geq R_{opt}^s$.  This result
in turn immediately implies the inequalities

\begin{eqnarray}
\label{RclipRbb}
R_{clip}^{s} & \leq & R_{bb} \\
\label{RstarsRB}
R_{*}^{s} & \leq & R_{B}\; ,
\end{eqnarray}
{\em with equality holding for both equations in the asymptotic limits
$\alpha\to\infty$ and $R_{G}\to 0$\/}.  This general behavior is
confirmed in fig.~\ref{Fig:compareOptTransfSpherical}, which shows the
results for the Gaussian scenario in the two relevant cases: zero and
non-zero bias.

Another available measure of the success of the optimal transformation
$\phi^{*}$ in rendering a good approximation for $\bm{J}_{B}$, is the
probability distribution $P(x_{*})$, where $x_{*}\equiv
\phi^{*}(x)/R_{*}^{s}$.  In order to obtain $P(x_{*})$, one just has
to recall that $P(x)$ is Gaussian with mean $R_{opt}^{s}$ and variance
$1-(R_{opt}^{s})^{2}$.  The optimal transformation is then
$x_{*}=\phi^{*}(x)/R_{*}^{s} =
\tanh(R_{opt}^{s}x/(1-(R_{opt}^{s})^{2}))/R_{*}^{s}$, and can be
regarded as an attempt to attach some structure to the distribution of
the transformed $x_{*}$.  With a simple change of variables,
$P(x_{*})$ is readily seen to be

\begin{equation}
\label{pxstar}
P(x_{*})  =  \frac{R_{*}^{s}\sqrt{1-(R_{opt}^{s})^{2}}}{\sqrt{2\pi}R_{opt}^{s}
(1-(R_{*}^{s}x_{*})^{2})} %\times
%\nonumber \\
% & \ & 
\exp
\left\{
\frac{-(1-(R_{opt}^{s})^{2})}{2(R_{opt}^{s})^{2}} \left[
\frac{1}{2}\ln\left(\frac{1+R_{*}^{s}x_{*}}{1-R_{*}^{s}x_{*}}\right)
-\frac{(R_{opt}^{s})^{2}}{1-(R_{opt}^{s})^{2}} \right]^{2}
\right\}\; .
\end{equation}
A comparison with eq.~\ref{px} shows that the two equations are very
similar, but not identical.  Some similarity in shape should indeed be
expected, mainly because $P(x_{*})$, just like $P_{CM}(x)$, {\em must
be\/} such that $(P(x)-P(-x))/(P(x)+P(-x))\propto x$, in order to
consistently prevent any further improvement by a similar
transformation.  One can verify in fig.~\ref{Fig:pxvspxprime} that the
resemblance between the probability distributions is closely
associated with the success of $R_{*}^{s}$ in saturating the upper
bound $R_{B}$.  The curves correspond to the Gaussian scenario with
$A=1/3$ and $B=0.1$ for two values of $\alpha$ (one can thus refer to
the upper solid curves of fig.~\ref{Fig:compareOptTransfSpherical}). 
Note that for $\alpha = 8$, the difference between $R_{B}$ and
$R_{*}^{s}$ is very small in fig.~\ref{Fig:compareOptTransfSpherical},
which is reflected in the solid curves of fig.~\ref{Fig:pxvspxprime}
being very close to each other.  Accordingly, the dashed curves in
fig.~\ref{Fig:pxvspxprime} get further apart for $\alpha = 10$ as the
mismatch between the overlaps increase in
fig.~\ref{Fig:compareOptTransfSpherical}.

\subsection{The biased case}

The simple biased case with $A=0$ and $B\neq 0$ provides an
interesting exception to the performances of $\bm{J}_{*}^{s}$ and
$\bm{J}_{clip}^{s}$.  The fact that $U(b)$ is linear implies $\partial
F/\partial R = 0$, as can be readily verified in
eq.~\ref{GL:gauss:sp:sp}.  This, on the other hand, implies the
equalities in eqs.~\ref{RclipRbb} and~\ref{RstarsRB}, that is,

\begin{eqnarray}
R_{clip}^{s} & = & R_{bb} \\
R_{*}^{s} & = & R_{B}\; .  
\end{eqnarray}
Therefore the strategy described in the previous section is successful
in attaining the opper bounds of section~\ref{Bayes}, and not only
asymptotically.  It should be noted that for a linear $U$, the vector
$\bm{J}_{opt}^{s}$ can be simply constructed with the Hebbian rule,
$\bm{J}_{opt}^{s}\propto\sum_{\mu}^{\alpha N}\bm{\xi}^{\mu}$,
$\forall\alpha$.  Therefore the best binary performance is attainable
by the clipped Hebbian vector $\bm{J}_{clip}^{s}$, in this case.  The
second equality, however, seems to us more remarkable, because it
stablishes a result which we could not find elsewhere in the
literature: {\em the optimal transformation manages to completely
incorporate the information about the binary nature of $\bm{B}$,
leading to the Bayes-optimal performance $R_{B}$ without the need of
explicitly constructing the center of mass of the Gibbs ensemble\/}. 
In other words, the technique of non-linearly transforming the
components of the vectors, introduced in~\cite{Bolle95a} and extended
in~\cite{Schietse95}, is able to give a definitive answer to the
problem it aims to solve.

\section{Conclusions}
\label{conclusions}

We have presented results on learning a binary preferential
direction from disordered data. Constraining the
candidate vectors to the binary space as well, we first
showed that Gibbs learning presents not only an 
exponential asymptotic decay and second order
``retarded learning'' phase transitions, but also first order phase
transitions for a simple Gaussian scenario.  

On the question of what is the optimal
estimator, given the data and the knowledge that the
preferential direction is binary, we have shown that the
answer depends on which space the estimator $\bm{J}$
is allowed to lie in. The best {\em continuous\/} estimator
is the Bayesian vector, which is the center of mass
of {\em binary\/} Gibbsian vectors. The best {\em binary\/}
vector is obtained by clipping the Bayesian
vector. We have calculated its properties in detail,
providing an upper bound to the performance of
binary vectors. 

Finally, we have also studied one possible way of
constructing approximations to these two
optimal estimators. By transforming the components of a
previously obtained continuous vector, we show that
the upper bounds cannot be saturated, in
general. Exceptions to this rule are the asymptotic
limits (both $R_{G}\to 0$ and $\alpha\to\infty$) and
the special case of a linear function
$U$. Interestingly, the linear
case also seems to be the only one in
which Gibbs sampling can be performed
without computational difficulties. We are therefore
left with a situation where the
approximations work perfectly only in the case where
they are actually not needed. We believe this kind of result
reinforces the need of investigating the
connection between results in statistical mechanics and computational
complexity theory.

\section*{Acknowledgements}

We would like to thank Manfred Opper for very fruitful
discussions. We also  acknowledge financial support from FWO 
Vlaanderen and the Belgian IUAP program (Prime Minister's Office). 
This work was supported in part by the Engineering Research Program 
of the Office of Basic Energy Sciences at the U.S. Department of
Energy under grant No. DE-FG03-86ER13606.

\section*{Appendix}

In order to show that eq.~\ref{Gamma} is correct, one just has to show
that the integral below vanishes identically, $\forall a$:

\begin{eqnarray}
\ & \ &\int{\cal D}z\, \left[ \mbox{sign}(az+a^{2})-|\tanh(az+a^{2})| 
\right] \nonumber \\
 & = &\int{\cal D}z\, \mbox{sign}(az+a^{2})\left[1 
-\tanh(az+a^{2})\right]  \nonumber \\
 & \stackrel{z = y -a}{=} &
\int \frac{dy}{\sqrt{2\pi}}e^{-(y-a)^{2}/2}\;
\mbox{sign}(ay)\; [ 1 - \tanh(ay) ] \nonumber \\
 & = & e^{-a^{2}/2}\int {\cal D}y\, \mbox{sign}(ay)\, e^{ay}
\left[ \frac{e^{-ay}}{\cosh(ay)} \right] \nonumber \\
 & = & 0\; .
\end{eqnarray}

%\bibliography{copelli}

\end{document}